\documentclass[11pt,twoside]{article}

\usepackage{asp2006}
\usepackage{epsf}
\usepackage{lscape}
\usepackage{graphicx}

\begin{document}
\title{Simulating the Formation and Evolution of Early-Type Galaxies:
Multi-phase treatment of the ISM, Star Formation and Feed-back}   
\author{Emiliano Merlin \& Cesare Chiosi}   
\affil{Department of Astronomy, University of Padova,
       Vicolo dell'Osservatorio 3, 35122 Padova, Italy}    

\begin{abstract}
We present the preliminary results of a "two-phases" description of
the ISM suited to NB-TSPH models of   galaxy formation and
evolution.
\end{abstract}


\noindent \textbf{Introduction}. The interstellar medium (ISM) of a
galaxy is made by at least four, nearly independent,  gaseous
phases: (i) the very cold, dense and clumpy molecular clouds; (ii)
the warm, neutral gas surrounding the molecular clouds; (iii) the
hot fully ionized tenuous material; (iv) and finally the very hot
and rarefied coronal gas expelled by supernova (SN) explosions.
Furthermore, stars  are observed to preferentially form inside the
molecular component. The four phases cannot be properly described by
the standard Smooth Particle Hydrodynamics (SPH), which is designed
to deal with a one-phase medium (Marri \& White 2003), and even more
important the whole gas content cannot be used to evaluate the star
formation (SF) rate. To model this complicated situation with the
Padua NB-TSPH code of galaxy formation and evolution (Carraro et al.
1998; Merlin \& Chiosi, 2006 and references) we consider (and
suitably modify) the \textit{sticky particles} algorithm of Levinson
\& Roberts (1981) in which the "cold particles" are meant to
represent self-gravitating clouds of molecular and neutral hydrogen,
i.e. two of the above phases lumped together. In brief, a hot gas
particle is turned into a cold gas particle if it is cooler than a
threshold temperature, denser than a threshold density, belonging to
a convergent flow, and losing thermal energy. When a gas particle
becomes cold, it immediately loses its SPH properties. Therefore  it
can freely  move across the tenuous ISM without feeling drag forces.
A cold particle is supposed to be made of two components of
different temperature: the "very cold" part corresponding to the
molecular core and the surrounding warm one. The  evolution of the
two components by thermal instability is conceived in such a way
that as the warm part cools down the mass of the "very cold" one
increases so that  their relative densities change (at given total
volume of the particle). A "very cold" particle is thus formed. This
is similar to what has been suggested  by Springel \& Hernquist
(2003) in cosmological context. The evolution of the "very cold"
particles is then governed only by gravity, radiative cooling,
cloud-cloud collisions (which dissipate kinetic energy), and SN
feed-back. A "very cold"  gas particle is eventually turned into a
star particle according to the Schmidt (1959) law, however
interpreted in the probabilistic manner proposed by  Lia et al.
(2002). Star particles can later be turned back into gas, as a
result of SN explosions. SNs release energy and metals to the ISM,
leading to the evaporation of the nearby clouds and thus
self-regulating the SF process.

\begin{figure} [!ht]
\centering
\includegraphics[width=12cm,height=3.5cm]{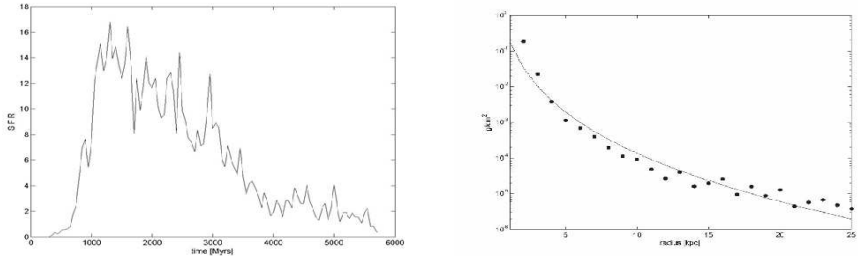}
\caption{Left: SF rate vs time. Right: Surface density profile of
star particles} \label{fig}
\end{figure}

\noindent \textbf{Results}.
 The above model has been used to simulate the formation and
 evolution of a galaxy whose initial conditions are derived from
cosmological density perturbations  according to  GRAFIC2 of
Bertshinger (2001) (see Merlin \& Chiosi 2006 for details). The
proto-galaxy has total mass (DM+BM) of $\sim 2 \times 10^{11}
M_{\odot}$ and  radius of 20 kpc. Each component is described by
$\sim$ 7000 particles. The BM is initially in form of gas. The
cosmological scenario is the W-Map3 $\Lambda$CDM. The proto-galaxy
is framed into the conformal Hubble flow, and it is followed from
the initial redshift z $\sim$ 60 down to z $\sim$ 1. Baryons follow
the DM perturbations until they heat up by mechanical friction,
radiative cooling becomes really efficient, the first cold clouds
begin to form, and eventually stars are born. Owing to the mass
resolution, each star particle has the mass size of  a star cluster,
in which real stars distribute according to a given initial mass
function. In a star particle SN explosions may eventually occur.
They release energy thus causing the evaporation of nearby clouds
and eventually quenching SF. Two important features of the model are
shown in Fig.\ref{fig}. The left panel displays  the SF rate
($M_{\odot}$ per year) vs age,  the right panel the spherically
averaged star surface density profile at z $\sim$ 1 (dots). The SF
consists of a single, prominent episode  at high redshifts followed
by a long tail. The formation of the galaxy is completed at z $\sim$
1.5. The mean metallicity is about $\sim$ 75$\%$ solar. Likely, the
stellar component has already relaxed to the Sersic (1968) profile
with $m \sim 4.3$ (solid line). Finally, the hot gas, heated by SN
explosions, causes strong galactic winds.

\noindent \textbf{Discussion}. Owing to the early, intense burst of
SF, the present model  may reproduce many features of the so-called
EROs (see e.g. Bundy et al. 2005). Fine tuning of the  model
parameters may yield results that closely fit  many observational
properties of early-type galaxies.



\end{document}